\documentclass[reprint,superscriptaddress,amsmath,amssymb,aps,prb,]{revtex4-2}
\usepackage{graphicx}
\usepackage{dcolumn}
\usepackage{bm}
\usepackage{color}
\usepackage[version=4]{mhchem}
\usepackage[colorlinks=True, linkcolor=blue, filecolor=magenta, urlcolor=blue,citecolor=blue]{hyperref}
\usepackage{mathcomp}
\usepackage{orcidlink}
\usepackage{upgreek}
\usepackage{pgffor}
\usepackage{eso-pic}

\newcommand{\rmc}{\mathrm{c}}

\begin{document}
\title{\textbf{Identifying Contributions to the Superconducting Diode Effect through Nonreciprocal Superconducting Properties}} 

\author{Ryo Iijima}
 \affiliation{Institute for Chemical Research, Kyoto University, Uji, Kyoto 611-0011, Japan}

\author{Hisakazu Matsuki}
 \email{matsuki.hisakazu.2i@kyoto-u.ac.jp}
 \affiliation{Institute for Chemical Research, Kyoto University, Uji, Kyoto 611-0011, Japan}
 \affiliation{Center for Spintronics Research Network, Institute for Chemical Research, Kyoto University, Uji, Kyoto 611-0011, Japan}

\author{Fugo Tokoro}
 \affiliation{Institute for Chemical Research, Kyoto University, Uji, Kyoto 611-0011, Japan}

\author{Susumu Hirata}
 \affiliation{Institute for Chemical Research, Kyoto University, Uji, Kyoto 611-0011, Japan}

\author{Akito Daido}
 \affiliation{Department of Physics, Graduate School of Science, Kyoto University, Kyoto 606-8502, Japan}

 \author{Ryusuke~Hisatomi}
 \affiliation{Institute for Chemical Research, Kyoto University, Uji, Kyoto 611-0011, Japan}
 \affiliation{Center for Spintronics Research Network, Institute for Chemical Research, Kyoto University, Uji, Kyoto 611-0011, Japan}

\author{Yoichi Shiota}
 \affiliation{Institute for Chemical Research, Kyoto University, Uji, Kyoto 611-0011, Japan}
 \affiliation{Center for Spintronics Research Network, Institute for Chemical Research, Kyoto University, Uji, Kyoto 611-0011, Japan}

\author{Shutaro Karube}
 \affiliation{Institute for Chemical Research, Kyoto University, Uji, Kyoto 611-0011, Japan}
 \affiliation{Center for Spintronics Research Network, Institute for Chemical Research, Kyoto University, Uji, Kyoto 611-0011, Japan}

\author{Youichi Yanase}
 \affiliation{Department of Physics, Graduate School of Science, Kyoto University, Kyoto 606-8502, Japan}

\author{Teruo Ono}
 \email{ono@scl.kyoto-u.ac.jp}
 \affiliation{Institute for Chemical Research, Kyoto University, Uji, Kyoto 611-0011, Japan}
 \affiliation{Center for Spintronics Research Network, Institute for Chemical Research, Kyoto University, Uji, Kyoto 611-0011, Japan}
 \affiliation{International Center for Synchrotron Radiation Innovation Smart, Tohoku University, Sendai 980-8577, Japan}


\begin{abstract}
Nonreciprocal superconducting responses provide a sensitive probe of superconducting states in systems where inversion and time-reversal symmetries are broken.
However, in the superconducting diode effect (SDE), where the critical current differs for opposite current directions, nonreciprocity can also arise from vortex-induced contributions caused by minute magnetic field misalignment, complicating its interpretation.
Here, we investigate a noncentrosymmetric \ce{[Nb/V/Ta]_{40}} superlattice and demonstrate nonreciprocity in all three fundamental superconducting properties: the critical current, critical field, and superconducting transition temperature.
By measuring the angular dependence of the nonreciprocal critical field $\Delta B_{\rmc 2}$, we separate the contribution of misalignment-induced out-of-plane vortices from the in-plane response.
We identify a sharply enhanced SDE emerging near $B_{\rmc 2}$, accompanied by nonreciprocal responses in $T_{\rmc}$.
These observations exhibit phenomenology similar to that reported in layered superconductors, where an orbital-FFLO state with finite-momentum Cooper pairing has been reported.
Our results establish a protocol for assessing out-of-plane vortex contributions to the superconducting diode effect and demonstrate that nonreciprocal responses provide complementary probes of unconventional superconducting states.
\end{abstract}
\maketitle

\section{Introduction}
Superconductivity is characterized by three fundamental properties: the superconducting transition temperature $T_\rmc$, the critical field $B_{\rmc 2}$, and the critical current $J_\rmc$, all of which are determined by the underlying superconducting free energy landscape~\cite{tinkham_introduction_2004}.
When this free energy landscape becomes asymmetric with respect to the relevant order parameter, nonreciprocal superconducting properties can emerge.
In recent years, nonreciprocal superconducting properties have attracted increasing interest, with most studies focusing on the superconducting diode effect (SDE)~\cite{nadeem_superconducting_2023}, characterized by unequal critical currents under opposite current directions. 
The SDE has been observed in a wide range of systems, including noncentrosymmetric \ce{[Nb/V/Ta]_n} superlattices~\cite{ando_observation_2020, kawarazaki_magnetic-field-induced_2022, kawarazaki_rectification_2023, miyasaka_observation_2021}, superconductor/ferromagnet heterostructures~\cite{suri_non-reciprocity_2022,
gutfreund_direct_2023,hou_ubiquitous_2023,narita_field-free_2022,narita_magnetization_2023,tokoro_Pt_2025,li_field-free_2025-1}, two-dimensional superconductors~\cite{bauriedl_supercurrent_2022,wu_field-free_2022,nagata_field-free_2025,jiang_zero-field_2025,han_intrinsic_2026,hu_polarityreversible_2026,nagata_field-independent_2026,nagahama_two-dimensional_2025}, kagome superconductors~\cite{Le_superconducting_2024}, and Josephson junctions~\cite{davydova_universal_2022,
jeon_zero-field_2022,pal_josephson_2022,wu_field-free_2022,he_spin-related_2023,matsuo_Josephson_2023,ghosh_high-temperature_2024,kim_intrinsic_2024,jeon_interferometric_2026,ren_modulation_2026}.
Among these diverse platforms, artificial superlattices have attracted particular interest because their composition, spin-orbit coupling, and electronic band structure can be systematically engineered through layer design, providing a highly tunable platform for investigating nonreciprocal superconducting phenomena.

In conventional Bardeen-Cooper-Schrieffer (BCS) superconductors~\cite{bardeen_theory_1957}, Cooper pairs are formed between electrons with opposite spins and momenta.
When inversion and time-reversal symmetries are both broken, the superconducting free energy can become asymmetric with respect to the direction of supercurrent. This asymmetry can give rise to nonreciprocal superconducting responses such as the SDE~\cite{daido_intrinsic_2022,yuan_supercurrent_2022}.
Consequently, the SDE has been proposed as a sensitive probe of the intrinsic modification of the superconducting state itself, such as unconventional pairing states or transitions in the electronic phase.
However, sizable nonreciprocal critical currents can also arise from asymmetric vortex dynamics induced by out-of-plane magnetic field components even in otherwise conventional superconductors~\cite{morrison_controlled_1970, ilin_critical_2014, chahid_high-frequency_2023, suri_non-reciprocity_2022, hou_ubiquitous_2023, he_geometric_2025, gutfreund_direct_2023, li_field-free_2025-1, swartz_asymmetries_1967,swartz_surface_1965}.
Since vortex contributions are highly sensitive to even a minute out-of-plane magnetic field, identifying and isolating such a contribution is essential for utilizing nonreciprocal properties as reliable probes of the underlying superconducting state.
One effective approach is to systematically analyze multiple nonreciprocal quantities under precisely calibrated magnetic field orientations.

\begin{figure*}[t]
    \begin{center}
    \includegraphics[scale=0.92]{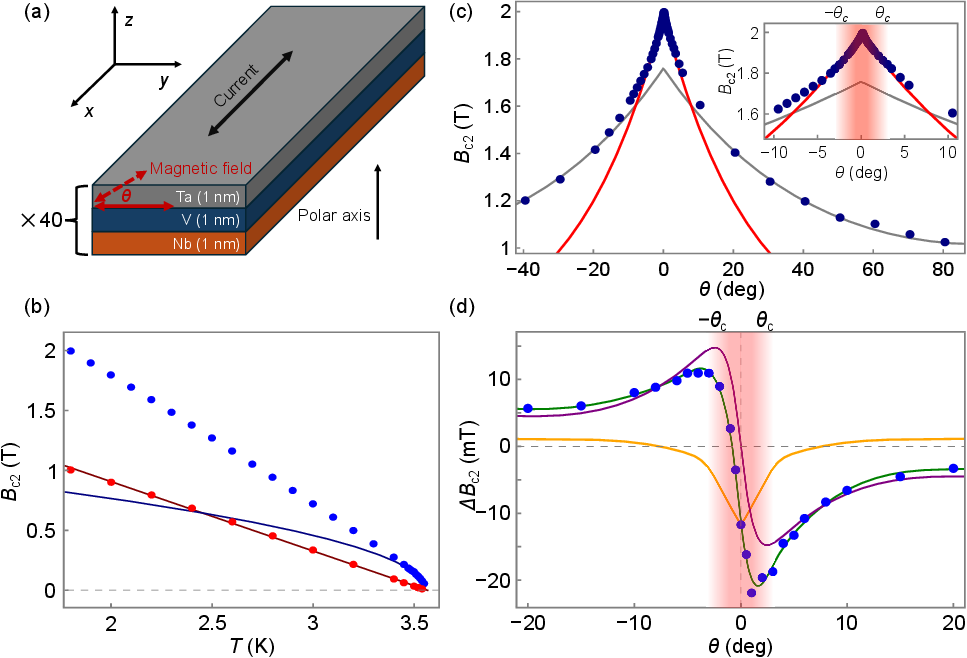}
    \caption{Device structure and superconducting properties of the [Nb(1~nm)/V(1~nm)/Ta(1~nm)]$_{40}$ superlattice. 
    (a) Schematic illustration of the $30~\mathrm{\upmu m}$-wide wire device and measurement configuration. The in-plane field is applied perpendicular to the current direction. In some measurements, the magnetic field is tilted out-of-plane by an angle $\theta$. 
    (b) Temperature dependence of the in-plane (blue) and out-of-plane (red) critical fields ($B_{\rmc 2 \parallel}$ and $B_{\rmc 2 \perp}$). The data are fitted using 2D and 3D Ginzburg–Landau (GL) models~\cite{tinkham_introduction_2004}, giving a $T_{\rmc}$ = 3.57 K at zero field.
    (c) Angular dependence of the critical field $B_{\rmc 2}(\theta)$ measured at 1.8 K with an applied current density of $J=556~\mathrm{A/cm^2}$.
    The red and gray curves are fits to the data extracted from the high-field region, $B_{\rmc 2}>1.8$~T, and the low-field region, $B_{\rmc 2}<1.5$~T, respectively, using a phenomenological expression combining the 2D and 3D Ginzburg--Landau angular forms~\cite{matsuoka_angle_2020} (see Supplemental Materials (SM) Eq.~3~\cite{supplement_2025}).
    The characteristic angle $\theta_\rmc$ is a guide to the eye indicating the angular range over which the critical field is sharply enhanced around the parallel alignment.
    (d) Angular dependence of the nonreciprocal critical field $\Delta B_{\rmc 2}(\theta)$ measured at 1.8 K under a large bias current $J = 0.95J_\rmc$.
    The nonreciprocal component of the critical field is defined as $\Delta B_{\rmc 2}=|B_{\rmc 2+}|-|B_{\rmc 2-}|$, where $B_{\rmc 2+}$ and $B_{\rmc 2-}$ denote the critical field magnitudes measured for positive and negative magnetic field directions, respectively, at a fixed current direction.
    The blue circles represent the experimental data. The green line is a smoothing spline fit to the data, which is decomposed into its symmetric (yellow line) and antisymmetric (purple line) components with respect to $\theta=0^\circ$ (see SM section II~\cite{supplement_2025}). These components correspond to the in-plane nonreciprocal response and the out-of-plane-field-induced vortex contribution, respectively.
    }
    \label{fig1}
    \end{center}
\end{figure*}

In this work, we show that the angular dependence of the nonreciprocal critical field, $\Delta B_{\rmc 2}$, in a \ce{[Nb/V/Ta]_{40}} artificial superlattice provides a direct means of identifying out-of-plane vortex contributions from the in-plane response.
By systematically varying the tilt angle of an otherwise in-plane magnetic field, vortex-induced nonreciprocal responses can be clearly identified.
Under the optimally aligned in-plane field, we identify a sharply enhanced SDE close to the upper critical field, accompanied by nonreciprocal responses in the superconducting transition temperature \(T_{\mathrm{c}}\). While the microscopic origin of this high-field nonreciprocity remains open, the observation of nonreciprocal behaviors in all three fundamental properties—\(J_{\mathrm{c}}\), \(B_{\mathrm{c}2}\), and \(T_{\mathrm{c}}\)—demonstrates that these parameters can serve as complementary and consistent probes for characterizing the high-field superconducting response. Furthermore, these high-field behaviors share qualitative features with theoretical models of layered superconductors under parallel magnetic fields, where unconventional regimes such as the orbital Fulde-Ferrell-Larkin-Ovchinnikov (FFLO) state have been proposed~\cite{fulde_superconductivity_1964,larkin_nonuniform_1964,nakamura_orbital_2024}. Our results demonstrate that a rigorous calibration of the field alignment is essential for evaluating nonreciprocal transport and establish that multiple nonreciprocal properties can serve as sensitive, complementary probes of high-field superconducting responses.

\begin{figure*}[t]
    \begin{center}
    \includegraphics[scale=0.9]{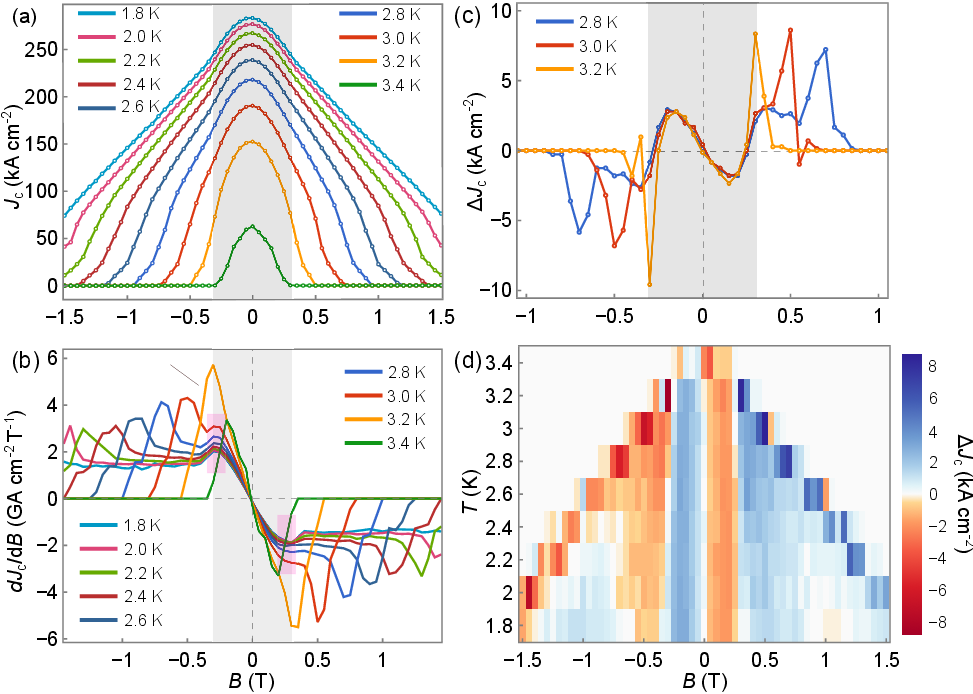}
    \caption{Critical current density--in-plane magnetic field ($J_{\rmc}-B$) characteristics and the superconducting diode effect. 
    (a) ($J_{\rmc+}-B$) phase diagram measured at different temperatures.
    $J_{\rmc+}$ indicates the current is applied along the $+x$ direction, while the magnetic field is applied along the in-plane +y direction ($B \parallel y$) ($J_{\rmc-}-B$ phase diagram is shown in SM~\cite{supplement_2025}.)
    (b) Differential critical current density ($dJ_{\rmc+}/dB$) as a function of increasing field measured at different temperatures. 
    An inflection point (the pink-shaded regions) is observed at $B$= $\pm$ 0.3 T for most temperatures well below $T_\rmc$.
    In addition, a secondary peak appears just below the superconducting phase boundary. 
    (c) Selected plots of the SDE measured at different temperatures. The nonreciprocal component of the critical current density is defined as $\Delta J_{\rmc} = |J_{\rmc+}| -  |J_{\rmc-}|$.
    A polarity reversal occurs near $B\approx0.3$ T, consistent with previous reports, while a distinct and sharp diode signal emerges close to the superconducting phase boundary.
    The gray-shaded regions in (a)--(c) indicate the low-field regime below $B = 0.3$ T. The boundary of this region coincides with the upturn in $J_{\rmc}$ and the polarity reversal of the SDE, suggesting that it may define a phase boundary. The complete set of field-dependent curves is shown in SM Fig.~S8~\cite{supplement_2025}.
    (d) Heatmap of the superconducting diode effect as a function of temperature and magnetic field. 
    }
    \label{fig2}
    \end{center}
\end{figure*}

\section{Results and Discussion}
A [Nb(1 nm)/V(1 nm)/Ta(1 nm)]$_{40}$ superlattice was grown on a MgO substrate by DC magnetron sputtering in an Ar atmosphere and patterned into a $30~\mathrm{\upmu m}$-wide wire by electron-beam lithography and Ar milling for electrical transport measurements, as shown in Fig.~1(a).
Unless otherwise noted, measurements were performed using a conventional four-terminal configuration with the magnetic field was applied in-plane and perpendicular to the current direction. 
In the following analysis, the critical current density $J_{\mathrm{c}}$, critical field $B_{\mathrm{c2}}$, and superconducting transition temperature $T_{\mathrm{c}}$ were determined using the criterion $R=R_\mathrm{N}/2$, where $R_\mathrm{N}$ is the normal state resistance.
While $B_{\mathrm{c2}}$ was obtained by linear interpolation, $J_{\mathrm{c}}$ and $T_{\mathrm{c}}$ were defined as the first measured points at which the resistance exceeded $R_{\mathrm{N}}/2$.

Figure 1(b) shows the temperature dependence of the in-plane (blue) and out-of-plane (red) critical fields. 
Near $T_\rmc$, the in-plane critical field $B_{\rmc 2\parallel}$ follows the two-dimensional Ginzburg-Landau (2D-GL) behavior, $B_{\rmc 2\parallel}\propto(1-T/T_{\rmc})^{1/2}$. Below $0.95T_\rmc$, it gradually deviates from this behavior and approaches an approximately linear temperature dependence.
From this analysis, we estimate a zero-temperature GL coherence length of $\xi_{\mathrm{GL}}(0) = 12.7~\mathrm{nm}$.
The layered nature of the superlattice becomes particularly evident in the angular dependence of the critical field at low temperatures. 
To accurately determine the in-plane field orientation, the out-of-plane tilt angle $\theta$ was calibrated by maximizing the critical current density, as shown in Supplemental Materials (SM) Fig.~S2~\cite{supplement_2025}, which includes Refs.~\cite{tinkham_introduction_2004, hou_ubiquitous_2023, matsuoka_angle_2020}.
As shown in Fig. 1(c), $B_{\mathrm{\rmc 2}}(\theta)$ at 1.8 K exhibits a pronounced cusp when the magnetic field is aligned parallel to the sample plane ($\theta \simeq 0^\circ$). 
This cusp is substantially sharper than predicted by the conventional Tinkham model for a two-dimensional superconductor~\cite{tinkham_introduction_2004} and becomes much less pronounced at higher temperatures (see SM Fig.~S1 ~\cite{supplement_2025}).
The sharp cusp observed in $B_{\mathrm{c}2}(\theta)$ near $\theta \simeq 0^\circ$ at low temperatures indicates that the superconductivity exhibits an enhanced two-dimensional confinement near the parallel alignment, which indicates the increase of two-dimensionality near the in-plane field orientation.
To characterize the angular range of this anomalous enhancement, we introduce a phenomenological angle scale, $\theta_\mathrm{c}$. Notably, as shown in Fig.~1(d), the symmetric component of the nonreciprocal critical field, $\Delta B_{\mathrm{c}2}(\theta)$, is also enhanced within a comparable angular range, pointing to a possible correlation between the critical field anisotropy and the near-in-plane nonreciprocity.

\begin{figure}[t]
    \begin{center}
    \includegraphics[scale=0.98]{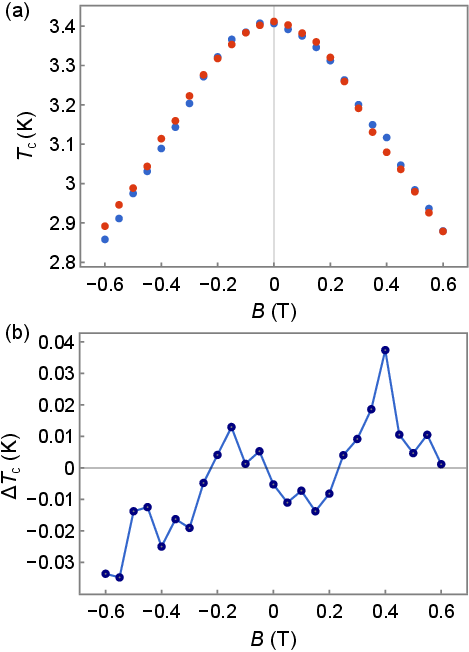}
    \caption{(a) $T_\rmc$, defined as the temperature at which $R = R_\mathrm{N}/2$, as a function of in-plane field.
    Blue (red) data indicate the applied current density $J=+ (-)55.6$ kAcm$^{-2}$, respectively.
    (b) The difference in $T_\rmc$ with current applied in positive and negative directions as a function of in-plane field. The multiple sign reversals were reproduced with $T_\rmc$ measurements.
    }
    \label{fig3}
    \end{center}
\end{figure}

To assess the possible contribution of field misalignment to the nonreciprocal response, we investigate the angular dependence of the nonreciprocal critical field $\Delta B_{\rmc 2}(\theta) = |B_{\rmc 2+}| - |B_{\rmc 2-}|$ as shown in Fig.~1(d). 
The raw angular dependence of the upper critical fields measured under positive and negative magnetic field directions, $B_{\text{c}2}^+(\theta)$ and $B_{\text{c}2}^-(\theta)$, are presented in SM Fig. S3~\cite{supplement_2025}.
The resulting \(\Delta B_{\mathrm{c2}}(\theta)\) is decomposed into symmetric and antisymmetric components with respect to $\theta=0^\circ$. The corresponding components calculated directly from the discrete experimental data are shown in SM Fig.~S5~\cite{supplement_2025}.
A contribution governed predominantly by the in-plane magnetic field is expected to be invariant under $\theta\rightarrow-\theta$ because reversing the tilt angle does not alter the magnitude of the in-plane magnetic field. It therefore contributes to the symmetric component of $\Delta B_{\rmc 2}(\theta)$.
In contrast, the out-of-plane field component changes sign under $\theta\rightarrow-\theta$, giving rise to an antisymmetric contribution associated with asymmetric vortex dynamics.
The symmetric component exhibits a pronounced peak at $\theta=0^\circ$ and is rapidly suppressed by the introduction of a minute out-of-plane magnetic field, demonstrating its extreme sensitivity to field alignment.
The angular width of this peak is comparable to the narrow angular range over which the anomalous enhancement of \(B_{\mathrm{c2}}(\theta)\) is observed in Fig. 1(c), suggesting a phenomenological correlation between the two features.
In contrast, the antisymmetric component reaches its maximum near $\theta\approx1^\circ$ and remains substantial over a much broader angular range. Reversing the current direction reverses the sign of the nonreciprocal response while preserving its overall angular dependence, as shown in SM Fig. S4~\cite{supplement_2025}. The qualitative separation into symmetric and antisymmetric components remains effective when the assumed zero-angle position is shifted by \(\pm0.2^\circ\), as shown in SM Fig. S6~\cite{supplement_2025}.

These results show that the nonreciprocal critical field response can be decomposed to odd and even functions of the tilt angle. This decomposition provides an experimental means of evaluating the sensitivity of the response to small out-of-plane field components and motivates the precise alignment procedure employed in the subsequent critical-current measurements.

Based on the angular criterion established from the critical field measurements, we next investigate the magnetic field and temperature dependence of the critical current density and the SDE, with the field aligned $+y$ within the sample plane. 
Figure 2(a) displays the positive-direction critical current density $J_{\mathrm{c}+}$ as a function of in-plane magnetic field at various temperatures. 
To emphasize changes in the field response, the corresponding differential critical current density, $dJ_{\mathrm{c}+}/dB$, is shown in Fig. 2(b). 
An inflection point in $J_{\mathrm{c}+}(B)$ is observed near $\pm$0.3~T across most of the measured temperature range, indicated by the gray shaded region in Fig.~2(a) and (b).
The same feature is also present for the opposite current polarity $J_{\mathrm{c}-}(B)$ (see SM Fig. S7~\cite{supplement_2025}). 
This reproducible anomaly defines a characteristic magnetic field scale separating two distinct regimes of critical current behavior.

We next examine the SDE, defined as $\Delta J_{\mathrm{c}}=|J_{\mathrm{c}+}|-|J_{\mathrm{c}-}|$, as shown in Fig.~2(c).
The full temperature and magnetic field dependence of the SDE is summarized in the heatmap in Fig.~2(d).
Two distinct contributions to the SDE can be identified: a broad low-field component below $\pm$0.4 T and a sharply enhanced high-field component that emerges immediately below the superconducting phase boundary.
Importantly, both components are measured under the nominally in-plane magnetic field and are physically distinct from the angle-antisymmetric, out-of-plane-vortex-induced contribution discussed in connection with Fig.~1(d).
The low-field component exhibits a nearly temperature-independent magnitude and undergoes a polarity reversal near 0.3 T at all temperatures below $T_\rmc$, consistent with previous reports on \ce{Nb/V/Ta} superlattices~\cite{kawarazaki_magnetic-field-induced_2022}.
Notably, this characteristic field scale coincides with the inflection point observed in $J_{\mathrm{c}\pm}(B)$ (gray shaded region in Fig.~2(b,c)), 
exhibiting a phenomenological correlation between the polarity reversal of the SDE and the change in the critical current-field dependence.

In contrast, a second SDE component emerges immediately below the superconducting phase boundary at high fields. 
This contribution appears as a narrow peak in $\Delta J_{\mathrm{c}}$, clearly separated from the broad low-field response~\cite{kawarazaki_magnetic-field-induced_2022}.
Remarkably, the high-field diode signal exhibits extreme angular sensitivity and is strongly suppressed by an out-of-plane field tilt of only $0.5^\circ$, corresponding to a perpendicular field component of only a few millitesla (see SM Fig. S9~\cite{supplement_2025}).

To understand the origin of these nonreciprocal responses, we consider alternative in-plane orbital mechanisms, such as superfluid density gradients~\cite{levichev_finite_2023} or asymmetric surface barriers for in-plane vortices. However, these mechanisms alone are difficult to reconcile with our observations. First, our $[\mathrm{Nb}(1~\mathrm{nm})/\mathrm{V}(1~\mathrm{nm})/\mathrm{Ta}(1~\mathrm{nm})]_{40}$ superlattice is periodic and lacks a macroscopic thickness gradient that would induce a superfluid density gradient as discussed for superconductor/normal metal bilayers~\cite{levichev_finite_2023}. Second, asymmetric surface barriers for in-plane vortices arising from the inequivalent vacuum and substrate interfaces may in principle generate a nonreciprocal critical current. In the vortex-rectification mechanism demonstrated by Hou et al.~\cite{hou_ubiquitous_2023}, the nonreciprocity originates from the combination of asymmetric vortex-entry barriers and Meissner screening currents, and becomes pronounced on an out-of-plane magnetic field scale of only a few Oe. In contrast, the characteristic features in the present Nb/V/Ta superlattice occur on substantially larger field scales, including the polarity reversal around 0.3 T and the additional enhancement close to the high-field superconducting phase boundary. In addition, the polarity reversal near 0.3 T and the sharp enhancement just below the superconducting phase boundary are not naturally expected within this picture, in which the sign of the diode effect is set by the field direction and the barrier asymmetry. Therefore, while vortex-related contributions cannot be excluded, the screening-current mechanism alone does not straightforwardly account for the characteristic field scales observed here.

To further examine the nonreciprocal properties from a complementary perspective, we investigated the current-direction dependence of $T_\rmc$. Figure~3 demonstrates that nonreciprocity equivalent to that observed in the critical current also appears in $T_\rmc$. For these measurements, the applied current density was fixed within the regime where the sharply enhanced high-field diode effect emerges. The difference in transition temperature between opposite current directions is defined as $\Delta T_{\mathrm c} = T_{\mathrm c+} - T_{\mathrm c-}$.

Remarkably, the field dependence of $\Delta T_{\mathrm c}$ closely follows that of the nonreciprocal critical current shown in Fig.~2(d), including multiple sign reversals. 
This correspondence indicates that $\Delta T_{\mathrm c}$ and $\Delta J_{\mathrm c}$ capture the same nonreciprocal response appearing in a certain region of the $(T,B,J)$ phase space, but from different measurement directions: $J_{\mathrm c}$ is obtained by sweeping current at fixed temperature and magnetic field, whereas $T_{\mathrm c}$ is obtained by sweeping temperature at fixed current and magnetic field. The disappearance of \(\Delta T_{\mathrm{c}}\) in the low-current limit, as shown in SM Fig. S10~\cite{supplement_2025}, is expected for a finite-current nonreciprocal response: as \(J \to 0\), the distinction between positive and negative current bias vanishes, and hence \(\Delta T_{\mathrm{c}} = T_{\mathrm{c}}(+J) - T_{\mathrm{c}}(-J) \to 0\). This behavior is consistent with the theoretical descriptions in which a finite supercurrent shifts the current-carrying superconducting state away from equilibrium and can render the superconducting phase boundary polarity dependent~\cite{yuan_supercurrent_2022}. Thus, the observed \(\Delta T_{\mathrm{c}}\) represents a nonreciprocal shift of the resistively determined superconducting transition under finite current bias rather than different equilibrium zero-current transition temperatures. Consequently, the $T_\rmc$ measurements provide a complementary view of the nonreciprocity observed in the critical-current measurements.

\section{Conclusion}
In conclusion, we have investigated nonreciprocal superconducting properties in a noncentrosymmetric Nb/V/Ta superlattice. By employing precise in-plane magnetic field alignment and analyzing the angular dependence of the nonreciprocal critical field, we distinguished the angular signature associated with minute out-of-plane field components from the nonreciprocal response that persists under nominally in-plane fields. This angular decomposition of $\Delta B_{\rmc 2}$ provides an experimental diagnostic for distinguishing field-misalignment-induced vortex contributions from nonreciprocal responses that persist under nominally in-plane fields.

Beyond the conventional superconducting diode effect, we observed correlated nonreciprocity in the critical current, critical field, and superconducting transition temperature. 
In particular, the observation of a nonreciprocal $T_\rmc$ demonstrates that nonreciprocity can be detected not only through critical current measurements but also through a complementary probe of the superconducting state.
The related field dependences, including multiple sign reversals, indicate that these observables probe a common region of the \((T,B,J)\) phase space from complementary measurement directions.
The strong angular sensitivity and characteristic field dependence of the high-field response are difficult to explain solely by out-of-plane vortex rectification. Moreover, the anomaly around 0.3 T, where the inflection point of \(J_{\mathrm c}(B)\) and the polarity reversal of the superconducting diode effect coincide, provides a characteristic field scale that is only weakly temperature dependent within the range where the feature is resolved. These observations are compatible with a change in the superconducting state and suggest a possible connection to finite-momentum pairing physics. However, the nature of the high-field state remains to be established, and vortex-related and other orbital mechanisms cannot be excluded.

An interesting direction for future work will be to clarify the origin of the substantially lower characteristic magnetic field scale observed in the present Nb/V/Ta superlattice compared with NbSe$_2$~\cite{wan_orbital_2023}.
Since the orbital limit of a thin-film superconductor can be understood by considering the formation of a vortex carrying one flux quantum, $(\pi/\sqrt{3})d\xi(T)B_{\rmc 2}^{||}\approx\Phi_0$, where $\Phi_0$ is the flux quantum, increasing the film thickness substantially reduces the orbital critical field. Given that the orbital-FFLO state shares key characteristics with such a vortex state, one potential explanation is the much larger structural periodicity of the superlattice (3 nm) relative to the interlayer spacing of NbSe$_2$ ($\sim$0.6 nm), which may alter the balance between interlayer coupling and orbital magnetic effects that govern finite-momentum pairing in layered superconductors. 
Systematic studies as a function of superlattice period and layer thickness could therefore provide direct insight into how the characteristic field scale of the nonreciprocal response is governed by the underlying layered structure.

\begin{acknowledgments}
This work was partly supported by the JSPS KAKENHI grant numbers JP20H05665, JP21K18145, JP24H00007, JP24H02233, JP25H01249,  JP26H00659, JP26H02016, JP26K17093, JST SPRING grant number JPMJSP2110, MEXT Initiative to Establish Next-generation Novel Integrated Circuits Centers (X-NICS) Grant Number JPJ011438, Collaborative Research Program of the Institute for Chemical Research, Kyoto University, and the Cooperative Research Project of RIEC, Tohoku University.

\end{acknowledgments}

\section*{Data Availability}
The data are available from the authors upon reasonable request.

\bibliographystyle{apsrev4-2}
%

\clearpage

\foreach \x in {1, 2, 3, 4, 5, 6, 7, 8, 9, 10, 11, 12} {
    \clearpage
    \thispagestyle{empty}
    \AddToShipoutPictureBG*{
        \AtPageLowerLeft{
            \includegraphics[page=\x, width=\paperwidth, height=\paperheight, keepaspectratio]{supplement.pdf}
        }
    }
    \mbox{} 
}

\end{document}